\documentclass{sig-alternate-10pt}
\usepackage[margin=1in]{geometry}

\usepackage{graphicx}
\usepackage{balance}  
\usepackage{afterpage}
\usepackage{pifont}
\usepackage{algorithmic}
\usepackage{amsmath}
\usepackage{listings}
\usepackage{empheq}
\usepackage[framemethod=tikz,xcolor=true]{mdframed}

\usepackage{amsthm}
\usepackage{amsfonts}
\usepackage{colortbl}
\usepackage{paralist}
\usepackage{enumitem}
\usepackage{epsfig}
\usepackage{forest}
\usepackage{graphicx}
\usepackage{multicol}
\usepackage{multirow}
\usepackage{pgf}
\usepackage{pdflscape}
\usetikzlibrary{positioning}
\usepackage{rotating}
\usepackage[normalem]{ulem}
\usepackage{relsize}
\usepackage{textcomp}
\usepackage{times}
\usepackage{tikz}
\usepackage{wrapfig}
 \usepackage{xspace}
\usepackage{balance}
\usepackage{thmtools,thm-restate}
\usepackage[normalem]{ulem}
\usepackage{url}
\usepackage{amsmath}
\usepackage[vlined,ruled]{algorithm2e}
\usepackage{xcolor}
\usepackage{mathtools}
\usepackage{subcaption}
\usepackage{xcolor}
\usepackage[utf8]{inputenc}

\usepackage[nocompress]{cite}
\usepackage{hyperref}
\usepackage{cleveref}

\newenvironment{denselist}{
    \begin{list}{\small{$\bullet$}}%
    {\setlength{\itemsep}{0ex} \setlength{\topsep}{0ex}
    \setlength{\parsep}{0pt} \setlength{\itemindent}{0pt}
    \setlength{\leftmargin}{0.8em}
    \setlength{\partopsep}{0pt}}}%
    {\end{list}}

\newcounter{enum}

\newcounter{tkwy}
\newcounter{chlng}

\newcommand\chlngcounter{\addtocounter{chlng}{1}\thechlng}
\newcommand\tkwycounter{\stepcounter{tkwy}\thetkwy}

\newcommand{\tar}[1]{\textcolor{orange}{[Tarique: #1]}}

\newcommand{\eat}[1]{}

\newcommand{\ww}[1]{\textcolor{blue}{Wentao: #1}}

\begin{document}
\pagestyle{empty}



\title{ML-Powered Index Tuning: An Overview of Recent Progress and Open Challenges}

\author{
    \alignauthor{
    Tarique Siddiqui
    \hspace{.8cm}
    Wentao Wu\\
    \vspace{.2cm}
           Microsoft Research\\
           \vspace{.2cm}
           \{tasidd, wentwu\}@microsoft.com
    }
}

\maketitle

\begin{abstract}
The scale and complexity of workloads in modern cloud services have brought into sharper focus a critical challenge in automated index tuning---the need to recommend high-quality indexes while maintaining index tuning scalability. This challenge is further compounded by the requirement for automated index implementations to introduce minimal query performance regressions in production deployments, representing a significant barrier to achieving scalability and full automation.
This paper directs attention to these challenges within automated index tuning and explores ways in which machine learning (ML) techniques provide new opportunities in their mitigation. 
In particular, we reflect on recent efforts in developing  ML techniques 
for workload selection, candidate index filtering, speeding up index configuration search, reducing the amount of query optimizer calls, and lowering the chances of performance regressions. 
We highlight the key takeaways from these efforts and underline the gaps that need to be closed for their effective functioning within the traditional index tuning framework. 
Additionally, we present a preliminary cross-platform design aimed at democratizing index tuning across multiple SQL-like systems—an imperative in today's continuously expanding data system landscape. We believe our findings will help provide context and impetus to the research and development efforts in automated index tuning.

\end{abstract}


\section{Introduction}
\vspace{1em}

Automated index tuning improves the performance of databases by recommending indexes that accelerate query execution. 
There has been extensive research over the past decades~\cite{FinkelsteinST88,KossmannHJS20}, and \emph{index tuners} have been developed for both commercial and open-source database systems~\cite{ChaudhuriN97,dta,dexter-1,ValentinZZLS00}.

\begin{figure}[t]
\centering
    \includegraphics[clip, trim=4cm 4cm 6.5cm 2.5cm, width=0.95\columnwidth]{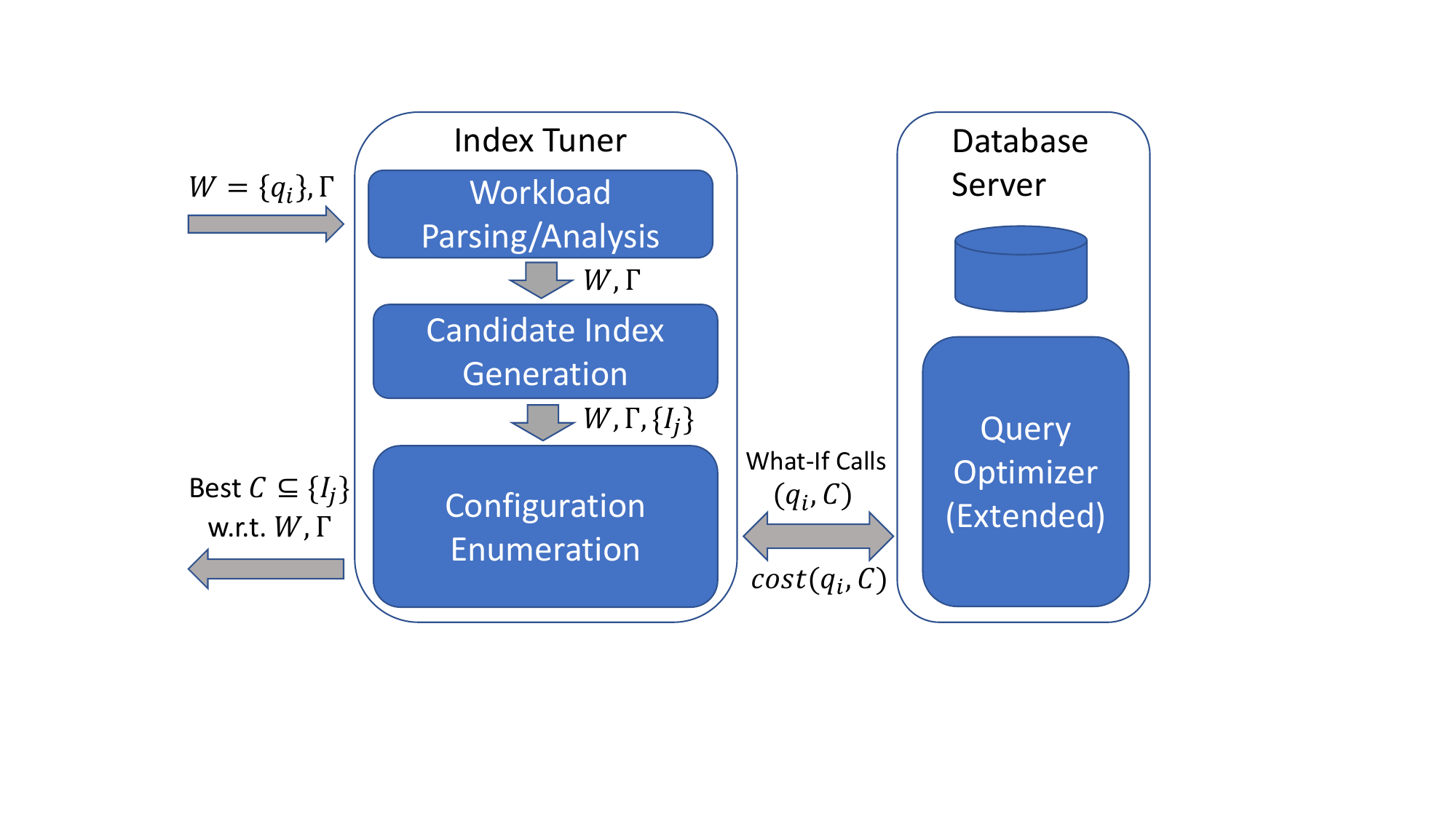}
\caption{The architecture of an index tuner, where $W$ 
is the input workload and $q_i\in W$ is a single SQL query, $\Gamma$ is a set of tuning constraints, $\{I_j\}$ is the set of candidate indexes generated for $W$, and $C\subseteq\{I_j\}$ represents an index configuration during enumeration.}
\label{fig:what-if-architecture}
\vspace{-1em}
\end{figure}

Figure~\ref{fig:what-if-architecture} presents the typical architecture of such an index tuner~\cite{ChaudhuriN97,ValentinZZLS00,dta}.
It contains three major components: (1) \emph{workload parsing/analysis}, where an input workload (of SQL queries) is parsed and analyzed; (2) \emph{candidate index generation}, which identifies a set of candidate indexes for each query in the workload; and (3) \emph{configuration enumeration}, which searches for an index configuration from the candidate indexes that meets the user-specified tuning constraints (e.g., the maximum number of indexes allowed or the total amount of storage taken by the indexes) while minimizing the total cost of the workload.\footnote{A configuration is defined as a set of indexes.}
For a configuration $C$ considered during enumeration, the index tuner leverages the \emph{what-if API}, an extended functionality of the query optimizer, to estimate the cost of each query on top of $C$ \emph{without} actually building the indexes contained by $C$~\cite{ChaudhuriN98}.
We refer to such query optimizer calls as ``what-if (optimizer) calls'' in this paper.
A what-if call can be time-consuming since it needs to invoke the query optimizer, especially for complex queries.

Despite this success, the recent advances in data management have exacerbated the existing challenges and posed new ones. 
We highlight three key problems.



\vspace{0.5em}
{ 
\em
\textbf{Problem \#1}: The growing scale and complexity of database SQL query workloads in modern cloud environments affect the quality of recommended indexes and contribute to increased time, cost, and resource overheads for index tuning.
}
\vspace{0.5em}

\begin{figure}	
	\hspace{-0.2cm}
	\centering
	\begin{subfigure}{0.49\linewidth}
		\centerline {
		\hbox{\resizebox{\columnwidth}{!}{\includegraphics{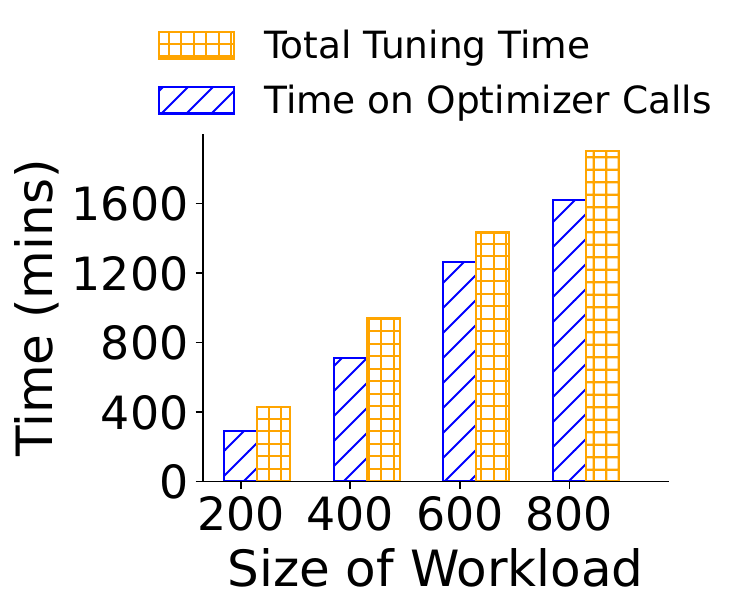}}}}
		\caption{Tuning time}
		\caption*{}
		\label{fig:motiv:tuning-time}
	\end{subfigure}
	\begin{subfigure}{0.42\linewidth}
		\centerline {
			\hbox{\resizebox{\columnwidth}{!}{\includegraphics{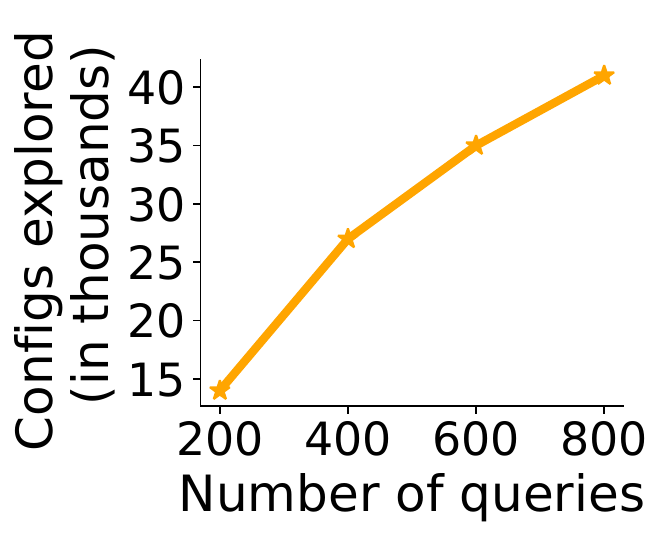}}}}
		\caption{\small Configurations explored}
		\label{fig:motiv:config}
	\end{subfigure}
	\caption{The growth in tuning time and configuration exploration on increasing workload size.}
        \vspace{-2em}
	\label{fig:motiv}	
\end{figure}

Cloud database services, such as Microsoft's Azure SQL Database~\cite{azure-sql}, host millions of databases with large and complex query workloads. Automatically and efficiently tuning indexes at that scale and complexity is a formidable task. In particular, the scalability of index tuning depends on (1) the number of queries in the workload, (2) the number of candidate indexes and resulting configurations that are enumerated, and (3) the number of optimizer invocations or what-if calls.
As depicted in Figure~\ref{fig:motiv}, we see that the tuning time for a state-of-the-art index advisor~\cite{dta} grows significantly as we increase the size of the workload. This is primarily because the space of configurations to explore increases (Figure~\ref{fig:motiv:config}), resulting in a large number of expensive what-if calls (consuming 70\% to 80\% of the overall tuning time).

\vspace{0.5em}
{ \em
\textbf{Problem \#2}: Minimal DBA monitoring and the potential impact on larger workloads in the cloud environments underscores the imperative to mitigate performance regressions stemming due to recommended indexes by index tuners.
}
\vspace{0.5em}

A major impediment to the goal of full automation and scalability is the requirement that index implementations should not cause significant query performance regressions~\cite{DasGIJJNRSXC19}. One important reason for query performance regression (QPR) is that index tuners use query optimizer's cost model (via what-if calls) to measure the improvement in query performance (e.g., execution time) due to recommended indexes~\cite{ChaudhuriN97,ChaudhuriN98,ValentinZZLS00}. While cost models are much more efficient than directly executing queries, they may not accurately capture the runtime behavior of queries, resulting in a mismatch between the actual and estimated query performance. The issue is further aggravated due to the scale, variety, and complexity of workloads, which make it hard to collect sufficient statistics or incorporate mechanisms for automatically identifying and fixing QPR~\cite{DasGIJJNRSXC19}.

\vspace{0.5em}
{ \em
\textbf{Problem \#3}: The current approach of building  system-specific and tightly-coupled index tuners is less tenable in today's fast-expanding landscape of rapidly growing number and variety of data systems. 
}
\vspace{0.5em}

\begin{figure*}
	\centerline {
		\hbox{\resizebox{0.92\textwidth}{!}{\includegraphics{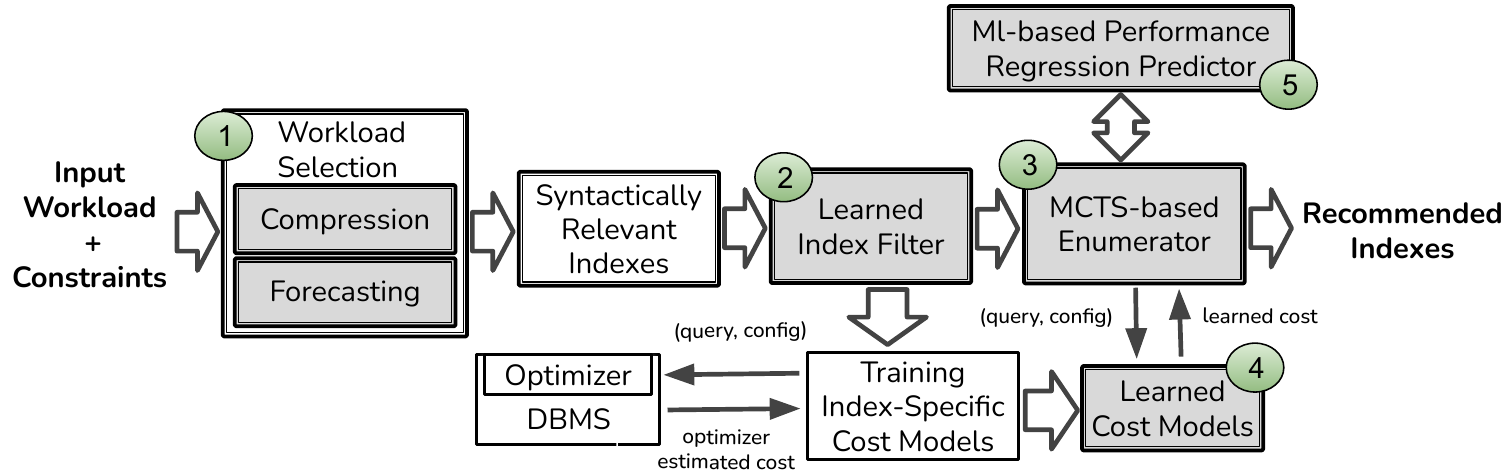}}}}
	\caption{ML-powered techniques (shaded) for improving index tuning.}
        \vspace{-1em}
	\label{fig:existingarch}
\end{figure*}
Modern enterprises manage many fast-evolving data systems, each optimized for different use-cases, and frequently add new ones. Data could reside in a variety of locations, e.g., operational stores, data warehouses, or data lakes~\cite{Polaris,Hyperspace,Helios}. Interestingly, only a limited number of database systems, such as Oracle, Microsoft SQL Server, IBM DB2, and PostgreSQL, support index tuning~\cite{ChaudhuriN97,dta,dexter-1,ValentinZZLS00}. This is surprising given that the process of index tuning is largely system-independent, with core components such as candidate index generation and configuration search algorithms reusable across systems with minimal changes. Yet, index tuners today are tightly coupled with specific database systems, and developing an index tuner for a new or evolving database system requires massive engineering efforts.


\subsection{Paper Overview}
\vspace{1em}

In this paper, we reflect on the recent efforts towards addressing the above challenges. While improving the  scalability of index tuning and addressing query performance regressions are not new problems, the  recent focus has largely been towards leveraging ML-powered techniques that can \emph{efficiently} identify useful configurations without sacrificing the quality of recommendations.
Another notable difference compared to prior work is that ML techniques require minimal changes to the underlying query optimizer or to the database system, and can potentially be integrated as ``bolt-on'' component(s) within existing time-tested and commercially deployed index tuning architectures~\cite{dta}.

\vspace{0.5em}
{ \em
   \textbf{Opportunity:} ML-powered techniques have the potential to interoperate with core index tuning components to improve the scalability and reduce query performance regressions, without significant changes to the index tuning architecture, the query optimizer, or the database system. 
}
\vspace{0.5em}


Figure~\ref{fig:existingarch} outlines an enhanced version of the index tuning architecture depicted in Figure~\ref{fig:what-if-architecture} after incorporating ML-based data-driven techniques.
It introduces novel software components and functionalities that improve the performance of the end-to-end index tuning workflow: (1) \emph{workload selection} that aims to reduce the size, complexity, and relevance of the input workload; (2) \emph{learned index filter} that aims to prune spurious candidate indexes with little impact on query performance; (3) \emph{MCTS-based enumerator} that aims to improve the effectiveness of index configuration enumeration; (4) \emph{learned cost models} that aim to reduce the number of what-if calls; and (5) \emph{ML-based performance regression predictor} that aims to reduce the chance of query performance regression.
We provide an overview of these new functionalities below, and the rest of this paper covers more details of each functionality as well as discussions on the opportunities and open challenges based on lessons learnt from our own experiences.

\vspace{-0.5em}
\paragraph*{Workload Selection}
We focus on two complementary sub-problems of \emph{workload compression} and \emph{workload forecasting}. Workload compression selects a small subset of queries from a large input workload, tuning which has the potential to result in as high-quality recommendations as tuning the entire workload. Workload forecasting, on the other hand, predicts arrival rate of queries for \emph{just-in-time} recommendation of indexes, reducing the size of workload that needs to be tuned at given point as well as improving the relevance of recommended indexes for queries in the near future (Section~\ref{sec:workload-selection}).

\vspace{-0.5em}
\paragraph*{Learned Index Filter}
After selecting SQL queries for tuning, the index tuner parses and analyzes the queries to generate \emph{synthetically relevant indexes} based on \emph{indexable columns}~\cite{ChaudhuriN97} (e.g., columns that appear in filter and join predicates appearing in the \emph{where} clause, as well as columns that appear in the \emph{group-by} and \emph{order-by} clauses).
It then tries to identify candidate indexes from the syntactically relevant indexes.
Many of such candidate indexes are turned out to be \emph{spurious}, meaning that they have little impact on query performance and can be safely pruned.
A learned index filter is developed based on this observation (Section~\ref{sec:filtering-spurious-indexes}).

\vspace{-0.5em}
\paragraph*{MCTS-based Enumerator}
As mentioned, configuration enumeration aims to find the best configuration from the candidate indexes provided.
A classic approach to configuration enumeration is \emph{greedy search}~\cite{ChaudhuriN97}, which suffers from scalability problems when facing a large search space with many candidate indexes and queries.
The MCTS-based enumerator aims to improve the effectiveness of configuration enumeration in large search space by identifying configurations that show promise and potential early on.
It leverages reinforcement learning (RL) techniques internally (Section~\ref{sec:search:rl}).

\vspace{-0.5em}
\paragraph*{Learned Cost Models}
The what-if calls used by index tuner can be expensive, especially when facing large and complex workloads.
One important observation we made is that many queries and configurations explored during configuration enumeration are \emph{similar}.
This opens up the door of leveraging ML techniques to learn in-situ lightweight cost models for clusters of similar queries and configurations during configuration enumeration, despite the fact that learning a generic cost model is extremely challenging.
We can significantly reduce the number of what-if calls by delegating many of them to the cost models learned (Section~\ref{section:distill}).

\vspace{-0.5em}
\paragraph*{ML-based Performance Regression Predictor}
The what-if calls used by the index tuner rely on the query optimizer's estimated costs, which can be off from the actual query execution time and result in QPR. 
An ML-based QPR predictor trained on top of query execution data can forecast and therefore avoid QPR firsthand.
We highlight the challenges of addressing QPR for production systems, giving an overview of recent efforts and the unsolved challenges that remain open (Section~\ref{sec:performance-regression}).

\vspace{-0.5em}
\paragraph*{Cross-platform Index Tuner}
Finally, to democratize the ML-powered index tuning techniques over multiple systems, we discuss the problems with the hitherto approach of developing \emph{system-specific} index tuners in today's expanding data system landscape. Towards addressing this, we propose an architecture for a \emph{cross-platform} index tuner, along with abstractions that will allow (the same) index tuning technologies to simultaneously benefit many data systems (Section~\ref{sec:cross-platform}).


\subsection{Scope and Limitations}
\vspace{1em}

Our primary focus in this paper is on improving the classical \emph{offline} index tuning process as used in commercial tools (e.g.,~\cite{Whang85,ChaudhuriN97,dta,ValentinZZLS00,BrunoC05,DashPA11,dexter-1,SchlosserK019}), and the adapted versions of them have also been deployed in modern cloud database services~\cite{DasGIJJNRSXC19}. 
Notably, there has been significant research efforts on \emph{online} index tuning techniques~\cite{SattlerGS03,SchnaitterAMP06,SchnaitterAMP07,BrunoC06,BrunoC07,SchnaitterP12,MozafariGY15,BasuLCVYSB15,SadriGL20,PereraORB21,perera2023no}, where the index tuner can create/drop indexes \emph{on the fly} to handle workload and data drifts. However, perhaps due to the inherent complexity and variety that comes with dynamic, ad-hoc, and non-stationary workloads, a consensus has not yet been reached on critical open questions of online index tuning such as the architecture, the optimization problem formulation, the optimality guarantee of the recommended indexes, and the performance evaluation criteria. Consequently, to the best of our understanding, such techniques have yet to find substantial adoption in commercial systems.

Meanwhile, there is a line of recent efforts on using ML for holistic database (knob) tuning (e.g.,~\cite{wang2021udo,van2017automatic,Trummer22,LiZLG19,ZhangAWDJLSPG18,ZhangLZLXCXWCLR19,ZhuLGBMLSY17,AkenYBFZBP21}) that goes beyond the scope of index tuning and therefore this paper.
There is also lots of related work on using ML for improving other specific aspects or components of database systems, such as physical data layout (e.g.,~\cite{HilprechtBR20,YangCWGLMLKA20}), buffer pool size (e.g.,~\cite{TanZLCZZQSCZ19}), and query optimizer (e.g.,~\cite{TrummerWMMJA19,MarcusNMZAKPT19,MarcusNMTAK22,ZhangIM0GLFHPJ22,YuC0L22}), which we omit in this paper as well.

Moreover, there are common challenges faced by applying ML techniques to solving data management problems that are not restricted to index tuning per se. There has been recent work on addressing such general challenges, such as reducing the overhead of generating training data~\cite{VenturaKQM21} and dealing with data updates/drifts~\cite{KurmanjiT23}. An in-depth discussion on these issues is worthwhile but beyond the scope of this paper.

\section{Workload Selection}\label{sec:workload-selection}
\vspace{1em}

The focus of workload selection has been in two directions: 1) selecting queries to tune, referred to as \emph{workload compression}, and 2) knowing when a query will arrive, referred to as \emph{workload forecasting}. We discuss representative research efforts in each direction.

\subsection{Workload Compression}
\vspace{1em}

A key factor affecting the scalability of index tuning is the number of SQL queries in the workload. In a typical cloud database service, a workload can contain hundreds or even thousands of queries. Tuning such a large workload in a reasonable amount of time is challenging. It is therefore natural to ask whether index tuning can be sped up significantly by finding a \emph{substitute} workload of smaller size while qualitatively not degrading the result of the application. It is crucial that this compressed workload can be found \emph{efficiently}; otherwise, the very purpose of compression is negated.

Prior workload compression techniques based on sampling and clustering~\cite{ChaudhuriGN02,DeepGKNV20} often fail to effectively capture the similarity between queries and miss out less frequent queries that may lead to substantial improvement in performance due to indexes. Furthermore, real workloads have typically more variety in query structures, which makes identifying relevant queries more challenging. To address these issues, we have developed ISUM, an indexing-aware and efficient workload summarization technology~\cite{isum}. ISUM employs two main techniques to identify relevant queries.

\emph{Measuring Potential Improvement:} We develop a new technique to 
efficiently estimate the potential in performance improvement of a query due to indexes \emph{without} requiring optimizer calls, which are key scalability bottlenecks. Our idea is to leverage statistics such as table size, selectivity, and costs of queries while eschewing parts of query optimization unrelated to indexing, to estimate improvement so that it is \emph{highly correlated} with the optimizer estimated improvements.

\emph{Capturing Indexing-aware Similarity:} On selecting a query, it is also important to quantify the improvement in performance on unselected queries in the workload due to indexes from the selected query. We represent each query as a set of features (derived from \emph{indexable columns}~\cite{ChaudhuriN97})
such that two queries with similar features will likely
result in similar sets of indexes. We weigh the features using statistics to capture their relevance to indexes. For instance, features on larger tables are more important, and similarly, the importance of indexable columns can
vary depending on whether they occur as part of the filter or join
predicates. 
We can further leverage ML techniques to automatically derive the weights of the features based on table size, selectivity, and position of the columns. Our feature representation also allows us to quantify the similarity between queries with different structures.

Combining both techniques, we measure the improvement due to each query over the entire workload, and develop a \emph{linear-time} algorithm that selects queries in decreasing order of their estimated improvement.

\vspace{0.5em}
{ \em
   \textbf{Takeaway \#\tkwycounter}: A workload compression technique for scalable index tuning requires efficient estimation of (1) potential performance improvement due to indexes, and (2) indexing-aware similarity between queries, both using minimal optimizer calls (a key scalability bottleneck).
}
\vspace{-0.5em}

\begin{figure}
	\centerline {
	\hbox{\resizebox{0.5\columnwidth}{!}{\includegraphics{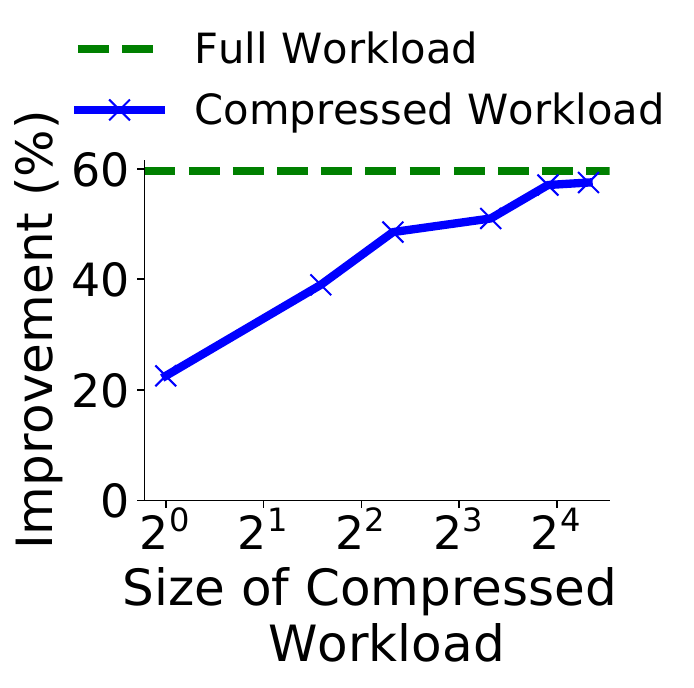}}}}
	\caption{Workload compression on TPC-DS.}
        \vspace{-1em}
\label{fig:promise}
\end{figure}

Figure~\ref{fig:promise} presents an example of running ISUM on the TPC-DS benchmark workload.
Overall, we observe that ISUM can lead to a median of 1.4$\times$
and a maximum of 2$\times$ performance improvements compared to prior techniques for the same compressed workload sizes. Furthermore, given an input workload consisting of queries along with
their costs, the time to select the compressed workload is small (<1\%) compared to the tuning time of the compressed workload~\cite{isum}. 

\eat{
\begin{figure}
	\centerline {
		\hbox{\resizebox{\columnwidth}{!}{\includegraphics{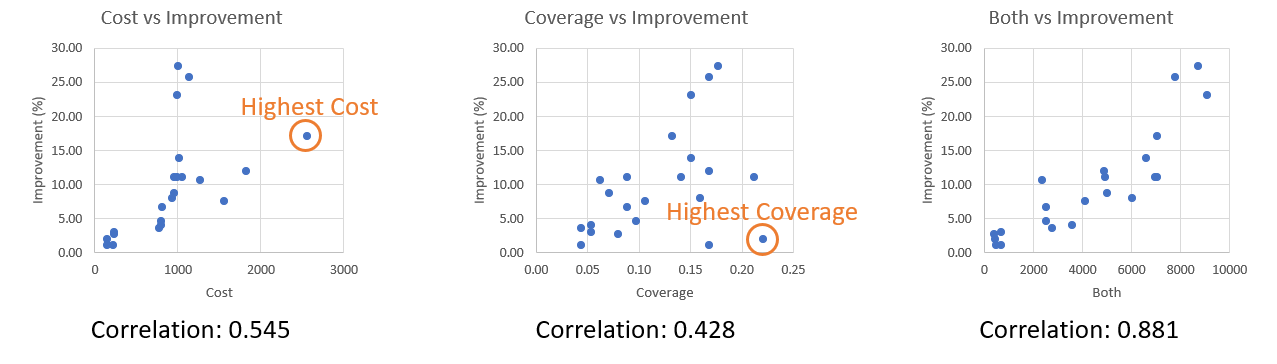}}}}
	\caption{Impact of cost and coverage as independent metric vs when combined together on improvement}
	\label{fig:costsimilarity}
\end{figure}
}

\eat{	
\begin{figure}	
	\hspace{-0.2cm}
	\centering
	\begin{subfigure}{0.23\textwidth}
		\centerline {
			\hbox{\resizebox{\columnwidth}{!}{\includegraphics{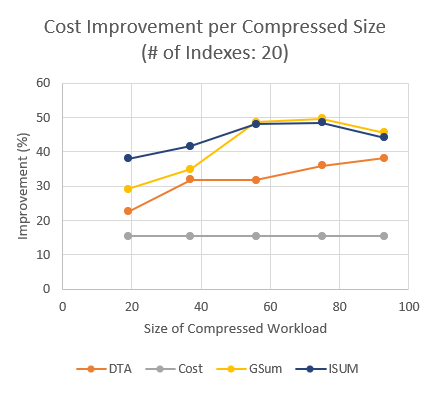}}}}
		\caption{TPC-DS (Improvement)}
		\label{fig:wctpcdsimprov}
	\end{subfigure}
	\begin{subfigure}{0.23\textwidth}
		\centerline {
			\hbox{\resizebox{\columnwidth}{!}{\includegraphics{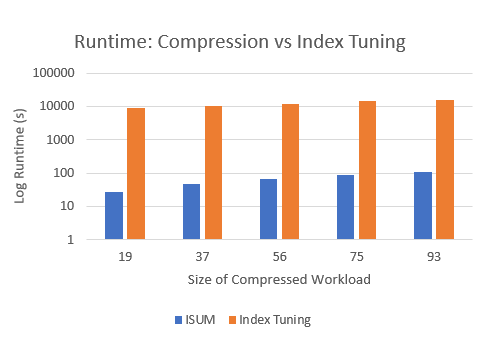}}}}
		\caption{TPC-DS (Runtime)}
		\label{fig:wctpcdsruntime}
	\end{subfigure}
	\caption{Improvement and runtime of compression algorithm \tar{need to replace}}
	\label{fig:wcresults}	
\end{figure}
}

\vspace{0.5em}
{ \em
   \textbf{Open Challenge \#\chlngcounter}:  The overhead associated with parsing of queries, collection of statistics, and measurement of improvements due to recommended indexes over the entire input workload can diminish the benefits of reduced tuning time from compression.
}
\vspace{0.5em}

Workload compression techniques, including ISUM, require that statistics such as selectivity, optimizer estimated cost of each query, and other physical plan characteristics are provided as input. 
We observe that most database systems expose functionality to collect such information.
For database systems where such information is not available, we need to make an optimizer call for each query in the workload, which is expensive for large input workloads.

\vspace{0.5em}
{ \em
   \textbf{Open Challenge \#\chlngcounter}: While existing workload compression techniques adopt specific optimization objectives, a more flexible workload characterization mechanism is needed for specifying ad-hoc constraints and user-guided selection of queries.
}
\vspace{0.5em}

Workload compression techniques use pre-defined criteria for selecting queries. However, in practice,  one may also want to obtain a representative subset with varying constraints, e.g., 100 most expensive queries while ensuring that every table in the database occurs in at least 5 queries, consuming at least a certain fraction of resources such as CPU and I/O. Thus, the specification for picking a representative subset of a workload depends on the task at hand and requires varying criteria and optimization goals. Additionally, it is crucial to characterize compressed workloads for interpretability. One direction to explore is to report the estimated improvement and drill-downs on how each query in the compressed workload represents queries in the workload that \emph{were not tuned}. Altogether, tighter integration of workload characterization mechanisms into a traditional index tuning engine and their evaluation for a broader set of tasks is an interesting area for future work.

\vspace{-0.5em}
\subsection{Workload Forecasting}\label{sec:workload-forecasting}
\vspace{1em}
Workload forecasting allows index tuners to make just-in-time recommendations for the workload expected to arrive in near future. Furthermore, workload forecasting can reduce the number of queries that index tuners need to analyze in each cycle. 

As one of the representative works, Ma et al.~\cite{ma2018query} develop a workload forecasting technique and leverage it to improve index tuning. It uses a two-phase framework. In the first phase, raw queries are pre-processed and clustered based on query templates (i.e., query instances without parameter binding). Clustering is necessary, as it is computationally infeasible to build models to capture and predict the arrival patterns for each template. In the next phase, an ML-based forecasting model is trained for each cluster that predicts how many queries the application will execute in the future (e.g., one hour from now, one day from now, etc.).

\vspace{0.5em}
{ \em
   \textbf{Takeaway \#\tkwycounter}: Predicting arrival rates of queries in near-future can help reuse traditional offline index tuners for scalable and just-in-time index selection.
}
\vspace{0.5em}

Workload forecasting partially mitigates the inability of offline index tuning in handling dynamic workloads (a core focus of online index tuning~\cite{BrunoC07}) while reusing the offline index tuners. The empirical findings show that when using forecasting, the throughput and latency of MySQL executing real workloads improve by 5$\times$ and 78\% over the 16-hour period when the indexes are added or removed after every hour. Similarly, over PostgreSQL, the technique achieves 180$\times$ better throughput and 99\% better latency~\cite{ma2018query}.

\vspace{0.5em}
{ \em
   \textbf{Open Challenge \#\chlngcounter}: A more holistic forecasting of future workloads, combining both arrival times as well as query instances (e.g., predicate values), is desired to enhance the quality of index recommendations.
}
\vspace{0.5em}

Prior work on index tuning as well as workload forecasting assumes that the query expressions remain unchanged over time. However, the recommended indexes may be sub-optimal when the expressions themselves evolve over time, e.g., a recurring analytical query that looks at last two days of sales data, or a query template that changes bindings based on the day the query runs or the same query template used by different teams with different parameter bindings.
Our analysis of enterprise workloads shows that while literal values may change over time, there are high-level patterns that can be learnt to predict the potential bindings in advance. Thus, an interesting direction for future work is to predict entire query instances in addition to the arrival times.

\section{Speeding up Index Tuning} 
\vspace{1em}

Searching for the best configuration in a large space with many candidate indexes is inherently challenging.
In fact, even a restricted version of the index selection problem is \emph{NP-hard}~\cite{Comer78} and/or even \emph{hard to approximate}~\cite{ChaudhuriDN04}.
State-of-the-art index search algorithms, such as the \emph{greedy} algorithm~\cite{ChaudhuriN97,dta,KossmannHJS20}, therefore rely on heuristics to reduce the search space. However, scalability and efficiency remain challenging even in such reduced search spaces. We discuss how we can take a data-driven perspective by leveraging ML techniques to speed up different components of index tuning. 

\subsection{Filtering Spurious Indexes}
\label{sec:filtering-spurious-indexes}
\vspace{1em}

\begin{figure}
	\centerline {
		\hbox{\resizebox{0.8\linewidth}{!}{\includegraphics{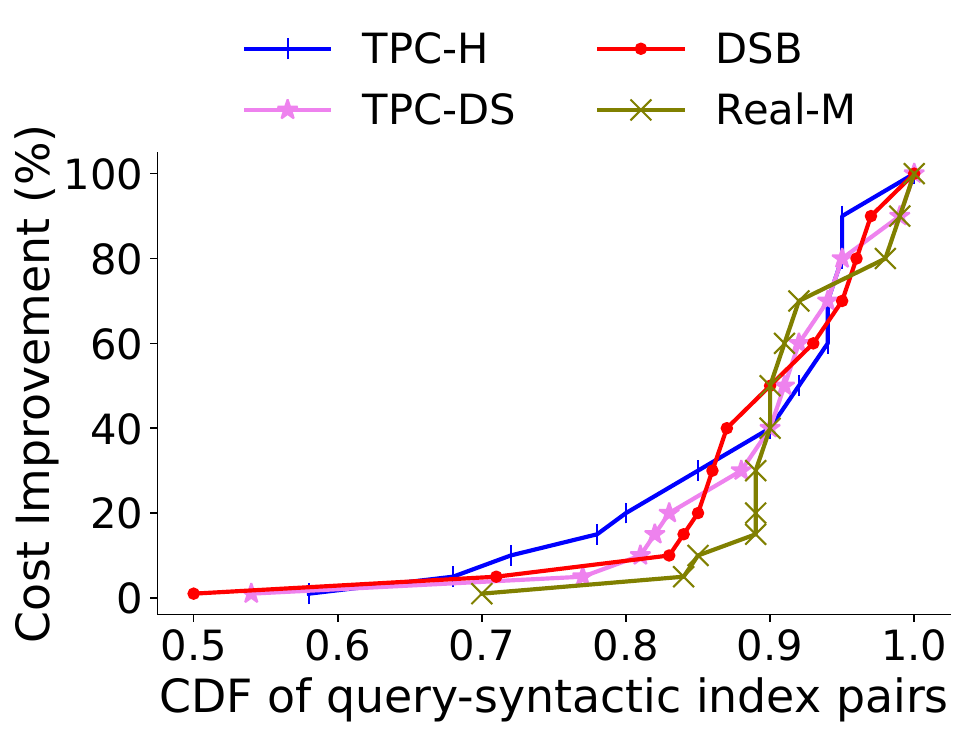}}}}
	\caption{Cost improvement for different fraction of query and syntactically relevant index pairs.}
	\label{fig:pruningopportunities}
 \vspace{-1em}
\end{figure}

Index tuners perform syntactic analysis (e.g., using a set of rules) to select an initial set of indexes for each query, called \emph{syntactically-relevant indexes}, for evaluation~\cite{ChaudhuriN97}. However, as showcased in Figure~\ref{fig:pruningopportunities}, we observe that 60\% to 70\% of such indexes are \emph{spurious}---they actually do not result in significant improvement in query performance~\cite{distillonger}. Thus, these spurious indexes can be filtered out and the optimizer calls made on these indexes can be avoided.

To prune such indexes early in the search process, we learn a \emph{workload-agnostic} model that uses structure and statistics information in the input (query, index) pair to identify when the index may not lead to a significant improvement in cost~\cite{distillonger}. 
We then use this model to remove a large number of spurious indexes. Our key insight is that we can probe the original physical plan of the query (i.e., the plan generated with existing indexes) to estimate the potential for improvement in the cost of the query due to a given index. For instance, if a join or sort operation is already efficient due to extensive filtering from earlier operations, adding an index that optimizes this operation is less beneficial. Similarly, if a filter column is not selective, we can easily prune an index that uses it as the leading key column. 
Furthermore, in many cases, we can compare the ordering of physical operators in the original plan
with the structure of the index to identify spurious indexes. Altogether, we capture many such signals and train a regression model to automatically learn rules to predict spurious indexes.

\vspace{0.5em}
{ \em
   \textbf{Takeaway \#\tkwycounter}:  Many syntactically-relevant indexes do not lead to improvement in performance. ML models trained on top of domain-specific signals can filter such spurious indexes in orders of magnitude less time compared to making what-if (optimizer) calls.
  }%
\vspace{0.5em}

\begin{figure}
	\centerline {
		\hbox{\resizebox{0.9\linewidth}{!}{\includegraphics{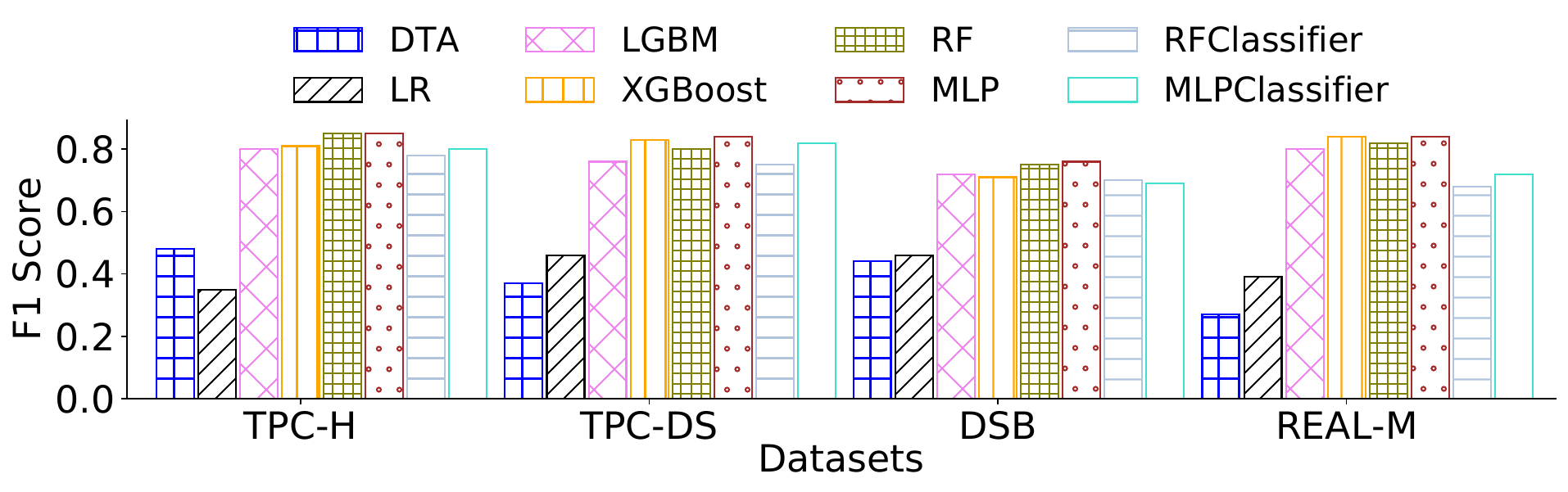}}}}
	\caption{Learned Index Filter.}
        \vspace{-1.5em}
	\label{fig:indexfilter}
\end{figure}

As shown in Figure~\ref{fig:indexfilter}, we find that index filtering models can be accurately learnt using (query, index) pairs generated from 3 to 4 databases and workloads and can remove over 70\% of the spurious indexes with a low rate (typically less than 10\%) of false negatives~\cite{distillonger}.

\eat{
\begin{figure}
	\centerline {
	\includegraphics[width=\columnwidth]{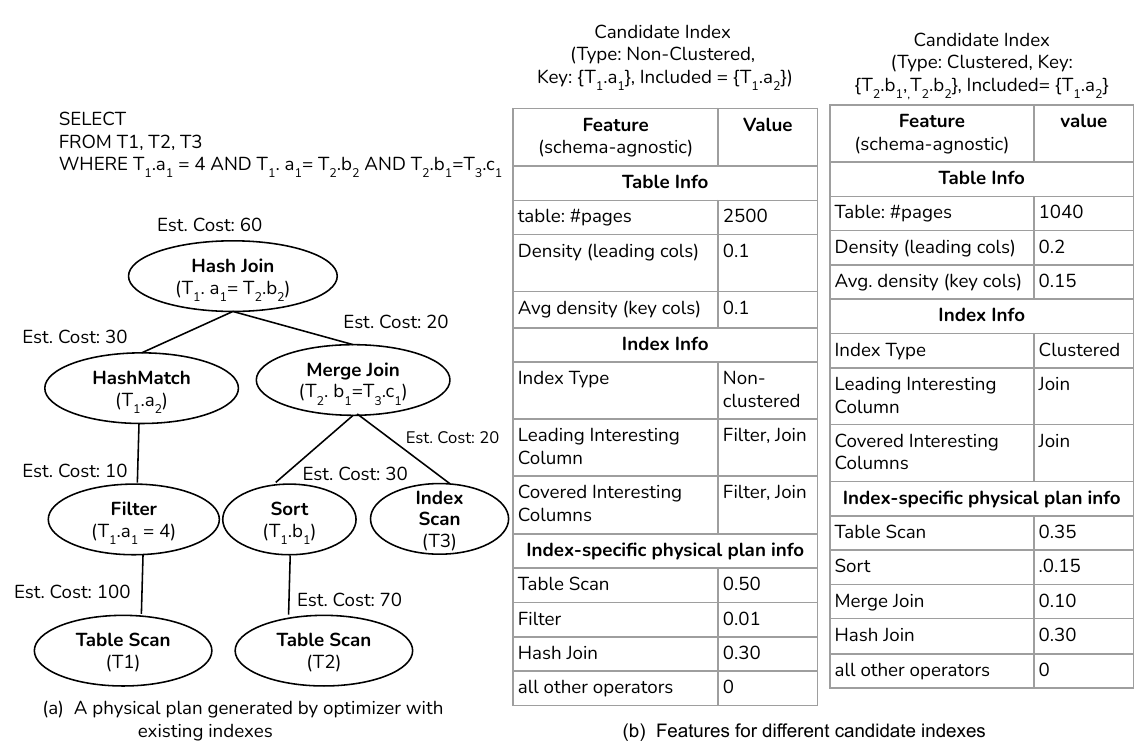}}
	\caption{Featurization for learned candidate filter}
	\label{fig:pruningmodel}
\end{figure}
}

\subsection{Search by Reinforcement Learning} 
\label{sec:search:rl}
\vspace{1em}

Given the large number of possible index configurations during configuration enumeration for cloud-scale workloads, it is practically impossible to have one what-if optimizer call for \emph{every} configuration and \emph{every} query enumerated.
This raises a trade-off between \emph{exploration} (of new configurations) and \emph{exploitation} (of promising configurations that are already known) when determining which configurations are worth what-if calls.
We develop a new index search framework based on Monte Carlo tree search (MCTS)~\cite{wu2022budget}, a classic reinforcement learning (RL) technology~\cite{BrownePWLCRTPSC12,sutton2018reinforcement}, to make better decisions on this exploration/exploitation trade-off.
In particular, we adapt the classic greedy search algorithm, typically used during configuration search~\cite{dta}, to handle the trade-off in a data-driven manner as follows:
\begin{denselist}
    \item \textbf{Exploitation:} We can expand configurations that \emph{show promise}, e.g., ones that contain the best configuration found by the greedy algorithm so far as a subset;
    \item \textbf{Exploration:} We can consider configurations that have been overlooked but may have \emph{potential} for improvement, e.g., ones that are not the \emph{winner} configuration found by the greedy algorithm, but have similar costs and can be utilized by more queries.
\end{denselist}


From this viewpoint, the existing greedy search approach can be viewed as one extreme---it relies on \emph{full} exploitation of what has been found with \emph{no} exploration.
Our RL-based approach, on the other hand, encourages more exploration, offering a principled way of tackling the above exploitation/exploration trade-off.

\vspace{0.5em}
{ \em
   \textbf{Takeaway \#\tkwycounter}:  RL-based techniques help navigate exploration and exploitation trade-offs more effectively on deciding which (query, configuration) to evaluate next.
  }%
\vspace{0.5em}
	
\begin{figure}
\centering
	\begin{subfigure}{0.9\columnwidth}
        \hbox{\resizebox{\columnwidth}{!}{\includegraphics{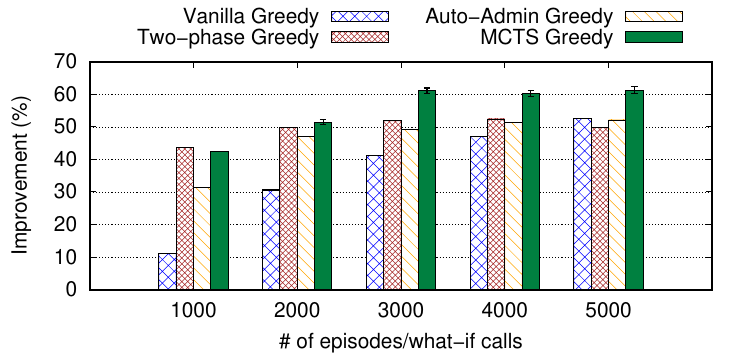}}}
		\caption{TPC-DS ($K=20$)}
		\label{fig:mcts:e2e:tpcds:k20}
	\end{subfigure}
	\begin{subfigure}{0.9\columnwidth}
        \hbox{\resizebox{\columnwidth}{!}{\includegraphics{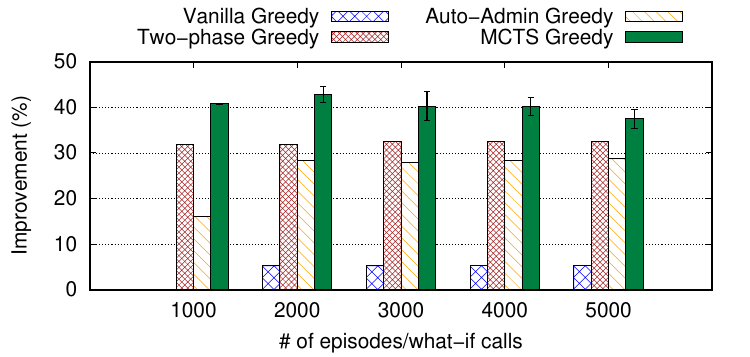}}}
		\caption{Real-M ($K=20$)}
		\label{fig:mcts:e2e:realm-large:k20}
	\end{subfigure}
\caption{Evaluation of MCTS configuration search.}
\vspace{-1em}
\label{fig:mcts:e2e}
\end{figure}

Figure~\ref{fig:mcts:e2e} presents evaluation results on the TPC-DS benchmark and a customer workload (Real-M) with the maximum desired configuration size $K$ set to 20. We compare the MCTS-based approach with both the \emph{vanilla} greedy search algorithm and its variants (shown as \emph{two-phase} greedy and~\emph{AutoAdmin} greedy in Figure~\ref{fig:mcts:e2e}) proposed in~\cite{ChaudhuriN97} and used in the Database Tuning Advisor (\emph{DTA}) developed for Microsoft SQL Server~\cite{dta}, which represents the current state of the art~\cite{KossmannHJS20}. As depicted in the figure, MCTS outperforms the greedy search algorithms consistently on both workloads w.r.t. the varying number of what-if calls. 

\vspace{0.5em}
{ \em
   \textbf{Open Challenge \#\chlngcounter}: Integrating MCTS-based search into commercial index tuning tools such as DTA remains an open problem, considering additional requirements such as anytime tuning, incremental handling of input workloads, and supporting reproducibility (difficult due to randomness inherent to MCTS).
   }%
\vspace{0.5em}

When the input workload is large and/or complex, we may want to run index tuning with a specific time-bound~\cite{dta}, or we may want to stop the tuning after some time without specifying a budget initially. Therefore, the search algorithm is desired to have the \emph{anytime} property, i.e., it should progressively find better configurations over time. This also requires incremental handling of more queries as input to the search algorithm and maintaining and reasoning about the intermediate state to minimize redundant work. Furthermore, the final recommended indexes can vary due to randomness in MCTS/RL, which affects reproducibility. Handling these challenges in a commercial tuning tool
like DTA requires non-trivial adaptations to the MCTS algorithm.

We note that there has been other recent work on ML-based configuration search~\cite{PereraORB21,abs-1801-05643,LanBP20,PereraORB22}, primarily targeting an online index tuning scenario. This line of work may be adaptable to offline index tuning but it shares the same challenges, as highlighted above, when it comes to integration with existing index tuners.

\subsection{Reducing What-If Optimizer Calls}
\label{section:distill}
\vspace{1em}


To achieve the best possible improvement in performance, the number of optimizer calls made during index configuration search can remain considerable despite pruning of spurious indexes and judicious enumeration of configurations.
To further improve the efficiency, we find that a significant number of optimizer calls for costing (query, configuration) pairs can potentially be replaced by more efficient data-driven cost models. 

Developing a general cost model that is independent of databases and workloads is hard due to the large varieties in the schema, query structures, and data distributions, 
despite the intensive efforts in the past decade (e.g.,~\cite{Ganapathi-berkeley09,AkdereCRUZ12-brown-icde,LiKNC12,WuCZTHN13,WuCHN13,WuWHN14,MarcusP19,SunL19,SiddiquiJQPL20,PaulCLS21,ZhaoCSM22}).
Our key observation to developing a lightweight cost model in the specific context of index tuning is that many queries in large workloads are \emph{self-similar}, e.g., multiple instances of the same stored procedure or query template parameterized differently. Many indexes explored during tuning can also be similar (e.g., sharing the same prefix of key columns, or influencing the same set of operators in the plan), which leads to similar configurations and results in similar cost reductions. As a result, the number of unique cost values is often much smaller than the number of index configurations explored during tuning (e.g., on average only 6 unique costs over 81 configurations explored per query for the TPC-H workload).

To leverage these characteristics, we group similar queries with the same query template and then learn a cost model for each group~\cite{distillonger}. For efficient in-situ learning during index configuration enumeration, we develop an \emph{iterative} training procedure (with optimality guarantees) and select diverse training instances (e.g., queries with different selectivities, indexes affecting different operators in the query, etc.) that minimize the number of optimizer calls for training each cost model (e.g., less than 50 optimizer calls per model on average across workloads). We show that it is possible to use low-overhead ML models that are significantly more efficient than making what-if optimizer calls. A key characteristic of these models is that they are \emph{agnostic} of the search algorithm (thus can be used by any algorithm), and do not require changes to the query optimizer.

\vspace{0.5em}
{ \em
   \textbf{Takeaway \#\tkwycounter}: ML-based cost models can be used as a generalized cache for similar (query, configuration) pairs, thereby avoiding many ``similar'' what-if calls.
  }%
\vspace{0.5em}

\begin{figure}
	\centerline {
		\hbox{\resizebox{0.8\linewidth}{!}{\includegraphics{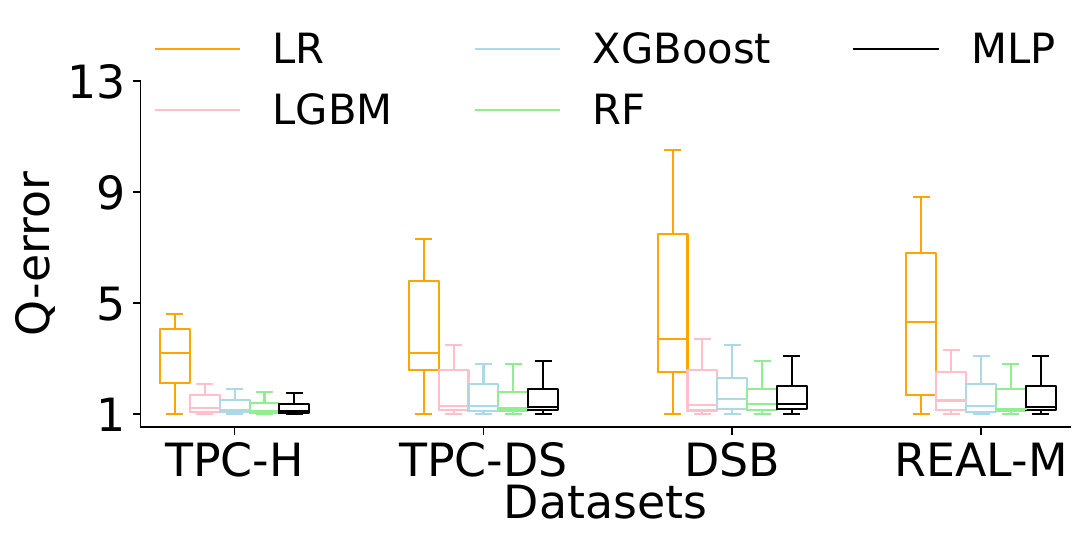}}}}
	\caption{Learned Index Cost Model.}
        \vspace{-1em}
	\label{fig:learnedcost}
\end{figure}

Figure~\ref{fig:learnedcost} depicts the effectiveness of different ML algorithms when used to train per-template cost models, with tree-based models achieving Q-error as low as 1.2. Furthermore, we find that combing ML-based cost models with filtering models for pruning spurious indexes (see Section~\ref{sec:filtering-spurious-indexes}) helps scale index tuning to large workloads without sacrificing the quality of the recommended indexes. For instance, for a TPC-DS workload with over 900 queries, combining index filtering and costing models can give index recommendations with similar quality as DTA but in $3\times$ less time.

\vspace{0.5em}
{ \em
   \textbf{Open Challenge \#\chlngcounter}: 
   Creating ML-based cost models across queries with different templates and across workloads can further improve the scalability by reducing training overhead (i.e., optimizer calls).
  }%
\vspace{0.5em}

The per-template cost models are less effective for workloads with many templates. There is a significant potential for reducing the number of optimizer calls as well as the number of models if we can generalize cost models across templates. There has been recent work on zero-shot cost models~\cite{hilprecht2022zero}; however, such techniques require a physical query plan (and thus an optimizer call) for featurization. 
Furthermore, in our current approach, we re-train during every tuning session from scratch due to limited mechanisms for meta-learning or fine-tuning learned models to capture workload and data drifts. 
This is another area where there are some intersections with online index tuning work~\cite{BasuLCVYSB15,SadriGL20,PereraORB21}.

\section{Performance Regression}
\label{sec:performance-regression}
\vspace{1em}

An important requirement of automated index tuning for production systems is creating
or dropping indexes should not cause significant query performance regressions (QPR), where a query’s execution cost increases dramatically after changing the indexes~\cite{DingDM0CN19}.
Such regressions are a major impediment for fully-automated and scalable index tuning~\cite{DasGIJJNRSXC19}. 
When an index tuner searches for optimal configurations, it compares
estimated improvements of query performance based on the optimizer's estimated costs.
Due to well-known limitations in the optimizer’s estimates,
such as errors in cardinality estimation~\cite{IoannidisC91} or cost modeling~\cite{WuCZTHN13}, using the
optimizer’s estimates can result in significant cost estimation errors.
The following trade-off is at the heart of why it is hard to achieve
scalability and low rate of QPR in index tuning simultaneously:


\vspace{0.5em}
{ \em
   \textbf{Efficiency vs. Accuracy Trade-off}: Optimizer's estimated costs
are much faster to compute, but they can be erroneous and result in low-quality recommendations and query performance regressions. On the other hand,
actual query execution time is much more accurate but it can only be obtained with significantly higher overhead, affecting the scalability of index tuning.
}
\vspace{0.5em}

One idea to reduce query performance regression is to selectively use execution time during index tuning along with optimizer's estimated cost. Towards this end, Ding et al.~\cite{DingDM0CN19} proposed a suite of ML techniques that learn over query execution telemetry collected from tens of databases to predict whether or not a new plan due to a selected index configuration has regressed with respect to another plan. Active learning techniques have been used to selectively collect query execution data for ML model training by deploying the same target database on non-production servers~\cite{ma2020active}.
Furthermore, techniques for fixing QPR have also been proposed~\cite{DingDM0CN19,DingDWCN18}.

\vspace{0.5em}
{ \em
   \textbf{Takeaway \#\tkwycounter}: Leveraging optimizer's estimated costs for index tuning but \emph{verifying} selected configurations at each step of configuration enumeration for query performance regressions via ML models trained over query execution statistics can reduce the chance of significant query performance regressions.
}%
\vspace{0.5em}

Unfortunately, from the scalability perspective, the inference process in~\cite{DingDM0CN19} is expensive since it requires query optimizer calls to obtain the physical plans of the queries. 
Indeed, the focus of~\cite{DingDM0CN19} was not scalability, targeting a closed-loop continuous tuning scenario where index tuning time is perhaps trivial considering the workload execution time, especially if there are query performance regressions.

\vspace{0.5em}
{ \em
   \textbf{Open Challenge \#\chlngcounter}: Efficiently detecting configurations causing query performance regressions without affecting the scalability of index tuning remains a challenge for large-scale workloads.
  }%
\vspace{0.5em}

A promising direction, intersecting with challenges discussed
in Section~\ref{section:distill}, is to learn pre-trained cost models that bridge the gap between optimizer cost models and the execution behaviour of
queries. A challenge that needs to be addressed is that such pre-trained models may not be accurate without requiring plan-level
details that need what-if optimizer calls. Toward this end, we can
explore techniques similar to the ones used for filtering spurious
indexes (Section~\ref{sec:filtering-spurious-indexes}) where the original physical plan is probed
with properties provided by an index to reason about potential
improvement in the cost, as showcased by the very recent work~\cite{ShiCL22}.
While learning a generalized model that
can work across workloads is challenging (as discussed earlier),
we can narrow down the problem by focusing only on indexing-specific
improvements. If we can accurately learn such models, it opens
up opportunities to eschew both optimizer calls as well as query
executions during index tuning, thereby significantly improving
the scalability.

\section{Cross-Platform Tuning}
\label{sec:cross-platform}
\vspace{1em}


The current database management landscape involves many SQL-like systems, with only a few supporting index tuning. While the systems differ in SQL dialects and functionalities (e.g., what-if API), the core ideas for index tuning can often be reused. This is more true for data-driven techniques discussed in this paper, where the ML models have limited dependency on the dialects or unique features of a particular system.

\vspace{0.5em}
{ \em
   \textbf{Open Challenge \#\chlngcounter}: There are many database systems (where indexes help improve performance) that either have no or low-quality automated index tuning capabilities, forcing users to manually select indexes for their workloads. Adding an index tuner to a new or evolving database system requires substantial engineering overhead, despite that many core ideas in index tuning are cross-platform reusable.
  }%
\vspace{0.5em}

\begin{figure}
	\centerline {
	\includegraphics[width=.95\columnwidth]{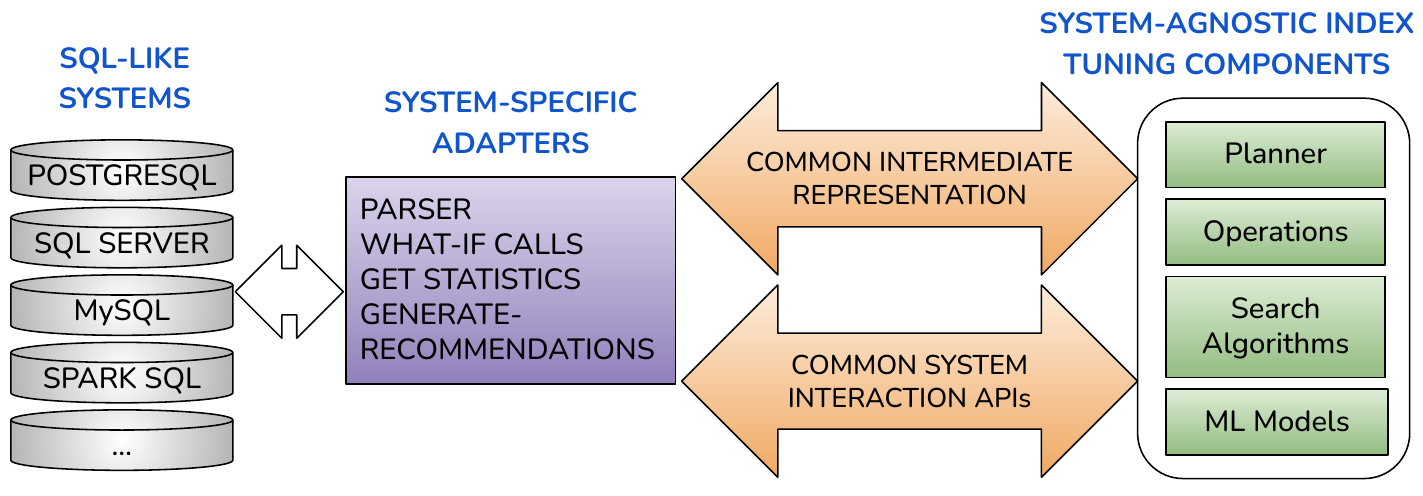}}
	\caption{A Cross-Platform Design for Index Tuning.}
        \vspace{-1em}
	\label{fig:onetuner}
\end{figure}

We hereby call for research efforts on developing a cross-platform index tuner that can work across multiple SQL-like systems, reusing core
index tuning techniques (e.g., data-driven ML models as well as
the search algorithms currently used in state-of-the-art index tuners). 
Similar efforts have been made in other areas such as query optimization~\cite{calcite,Magpie}.
A cross-platform index tuner needs to adapt to the heterogeneity of features varying across database engines, while reusing the
common steps as much as possible. We abstract such a system in
Figure~\ref{fig:onetuner} with the following main components:

\begin{itemize}
    \item \textit{Common Data Representation (IR)} consisting of a basic set of elements that we need to capture across systems, e.g., database, tables, columns, logical operators, physical operators, and sub-plans. A cross-language specification such as Subtrait~\cite{substrait} can potentially be leveraged for IR.
    \item \textit{Common System Interaction APIs} consisting of a common set of APIs that can be used to interact with the database during index tuning. Examples of such APIs include ones for query optimization in the presence of one or more indexes, query execution, creation of hypothetical indexes similar to the what-if API, and building of statistics.
    \item \textit{Adapters} providing the system-specific implementation of the common system interaction APIs that vary across systems.
    \item \textit{Index Tuning Planner} enabling cross-platform tuning functionality. It considers system-specific features and user requirements, and outputs an index tuning plan (analogous to query execution plan in database systems). The tuning task can be represented with a small set of index tuning \emph{operators}
    (e.g., \texttt{enumerate}, \texttt{combine}, \texttt{evaluate}) that can be composed together and configured to perform index tuning. The index tuning plan can be an \emph{acyclic graph} of these operators that is dynamically constructed and optimized by the planner based on system features and user requirements.
\end{itemize}

Overall, a cross-platform index tuner consisting of the components as envisioned above has the potential to democratize index
tuning to many more systems than those that are currently supported. In
addition, such an index tuner will allow (a) borrowing of the best concepts
(implemented as operators) from different index-tuning algorithms,
(b) independent improvement and maintenance of the functionality
of operators, and (c) extensibility by incorporating new techniques
(implemented as new operators) in the future without rewriting the
algorithms or changing unrelated operations.

\section{Conclusion}
\vspace{1em}
In this paper, we have highlighted the challenges inherent to automated index tuning, which are further exacerbated within modern cloud environments, and we have discussed recent efforts and opportunities in leveraging ML-powered techniques to address them. We presented an end-to-end analysis of the index tuning workflow, with a focus on the core components such as workload selection and configuration search. We described the issue of query performance regression (QPR) and discussed ML techniques for addressing QPR without affecting index tuning scalability. We also sketched the design of a cross-platform index tuner that extends the current index-tuning software stack to support multiple SQL-like systems. 
We believe this paper will help create awareness of recent progress and highlight
open challenges for future research in index tuning.

\vspace{-0.5em}
\paragraph*{Acknowledgement}
We thank Surajit Chaudhuri and Vivek Narasayya for their valuable feedback on this work.


\balance


\bibliographystyle{abbrv}
\bibliography{main}

\end{document}